\documentclass[traditabstract]{aa}
\def\gsim{\lower.4ex\hbox{$\;\buildrel >\over{\scriptstyle\sim}\;$}}
\def\lsim{\lower.4ex\hbox{$\;\buildrel <\over{\scriptstyle\sim}\;$}}
\renewcommand{\vec}[1]{\mbox{\boldmath $#1$}}
\def\Om{\Omega}
\def\q{\qquad}
\def\beg{\begin{eqnarray}}
\def\ende{\end{eqnarray}}
\def\gsim{\lower.4ex\hbox{$\;\buildrel >\over{\scriptstyle\sim}\;$}} 
\def\lsim{\lower.4ex\hbox{$\;\buildrel <\over{\scriptstyle\sim}\;$}}

\renewcommand{\textrm} [1] {\rm #1} 
\renewcommand{\textit} [1] {\it #1}
\renewcommand{\textbf} [1] {\bf #1} 
\usepackage{graphicx}
\usepackage{txfonts}
\begin{document}

\title{The angular momentum transport by standard MRI in quasi-Kepler cylindric Taylor-Couette flows}

\author{ M. Gellert \and G. R\"udiger \and  M. Schultz}

\institute{Leibniz-Institut f\"ur Astrophysik Potsdam,
           An der Sternwarte 16, D-14482 Potsdam, Germany\\
           \email{mgellert@aip.de}}

\date{}
%%%%%%%%%%%%%%%%%%%%%%%%%%%%%%%%%%%%%%%%%%%%%%%%%%%%%%%%%%%%%%%%%%%%%%%%
 \abstract 
  {The instability of a quasi-Kepler flow in dissipative Taylor-Couette systems under the presence of an homogeneous 
   axial magnetic field is considered with  focus to the excitation of  nonaxisymmetric modes and the resulting angular 
   momentum transport. 
   The excitation of nonaxisymmetric modes requires higher rotation rates than the excitation of the axisymmetric mode 
   and this the more the higher the azimuthal mode number $m$. We find that the weak-field branch in the instability map
   of the nonaxisymmetric modes has always a positive slope (in opposition to the axisymmetric modes) so that for given 
   magnetic field the modes with $m>0$ always have an {upper limit} of the supercritical Reynolds number. In order to 
   excite a nonaxisymmetric mode at 1 AU in a Kepler disk a minimum field strength of about 1 Gauss is necessary. For 
   weaker magnetic field the nonaxisymmetric modes decay. \\ 
   The angular momentum transport of the nonaxisymmetric modes is always positive and depends linearly on the Lundquist 
   number of the background field. The molecular viscosity and the basic rotation rate do not influence the related 
   $\alpha$-parameter. We did not find any indication that the MRI decays for small magnetic Prandtl number as found by 
   use of shearing-box codes. At 1 AU in a Kepler disk and a field strength of about 1 Gauss the $\alpha$ proves to be 
   (only) of order 0.005.
  }
  
\keywords{instabilities -- MHD}
               
\titlerunning{Standard MRI in quasi-Kepler Taylor-Couette flows}

\authorrunning{M. Gellert, G.~R\"udiger \& M. Schultz}

\maketitle

%%%%%%%%%%%%%%%%%%%%%%%%%%%%%%%%%%%%%%%%%%%%%%%%%%%%%%%%%%%%%%%%%%%%%%
\section{Introduction}
%%%%%%%%%%%%%%%%%%%%%%%%%%%%%%%%%%%%%%%%%%%%%%%%%%%%%%%%%%%%%%%%%%%%%%%
The longstanding problem of the generation of turbulence in various
hydrodynamically stable situations has found a solution in recent years
with the MHD shear flow instability, also called magnetorotational instability (MRI), in
which the presence of a uniform axial magnetic field has a destabilizing effect on a
differentially rotating flow with the angular velocity decreasing
outwards.

According to the Rayleigh criterion an ideal flow is stable against
axisymmetric perturbations whenever the specific angular momentum increases
outwards
\beg
\frac{{\rm{d}}}{{\rm{d}}R}(R^2\Om)^2 > 0,
\label{eq1}
\ende
where ($R$, $\phi$, $z$) are cylindrical coordinates, and $\Om$ is the
angular velocity.  Kepler rotation with $\Om\propto R^{-1.5}$ is thus hydrodynamically 
stable against axisymmetric perturbations. It is not stable, however, under the presence of a uniform axial 
magnetic field (MRI). It is also not stable against nonaxisymmetric perturbations under the presence of a 
toroidal field (AMRI). In these cases the stability condition (\ref{eq1}) simplifies to 
\beg
\frac{{\rm{d}\Om^2}}{{\rm{d}}R} > 0.
\label{eq2}
\ende
For real fluids the instability condition looks  more complicated. In order to be unstable the rotation must be 
fast enough and the magnetic fields must not be too weak or too strong. The lower magnetic  limit is fixed by the 
electric conductivity of the plasma while the geometry of the disk determines the upper magnetic limit.  As  
protoplanetary Kepler disks are so cool that even the action of MRI is in question (``dead zones"). In the present 
paper the lower limit is attacked with the calculations.  For numerical estimations we shall use the values 
\beg
\eta\simeq 4\cdot 10^{15}\ {\rm cm^2/s}, \ \ \ \ \ \ \ \  \ \ \ \ \ \ \ \ \ \ \ \ \ \ \ \    \rho\simeq 10^{-10}\ {\rm g/cm^3},
\label{eq3}
\ende
(see Brandenburg \& Subramanian 2005) for which we have derived a minimum field strength of 0.1 Gauss in order to excite 
unstable axisymmetric modes in the protoplanetary disk (R\"udiger \& Kitchatinov  2005). This value is based on the calculation 
of the minimum magnetic Reynolds number 
\beg
{\rm Rm'}=\frac{\Om H^2}{\eta}
\label{Rm}
\ende
for fixed values of the half-thickness $H$ of the disk and the magnetic Prandtl number 
\beg
{\rm Pm}=\frac{\nu}{\eta}.
\label{Pm}
\ende 
The vertical magnetic field $B_z$ is  measured in terms of the Lundquist number 
\beg
{\rm S'}=\frac{B_z H}{\sqrt{\mu_0\rho}\eta},
\label{S}
\ende
for which the ${\rm Rm}'$ is the absolute minimum for given ${\rm Pm}$. Note that in the definitions of ${\rm Rm}'$ and ${\rm S}'$ 
the molecular viscosity does not appear. The viscosity only  appears  in the definition of the magnetic Prandtl number (\ref{Pm}).

The instability map ${\rm Rm'=Rm'(S')}$ for the {\em axisymmetric} mode and for fixed Pm is very characteristic. 
For fixed Rm' exceeding a minimum value of order 10 the instability exists between  an upper  limit and a lower 
limit of ${\rm S}'$ which itself must exceed  a minimum value of order unity (see Fig. 2 in RK05). In the following 
the curve formed by the upper limits is called the strong-field branch and the curve formed by the lower limits is 
called the weak-field branch. Both the branches have opposite slopes. 
For given $\rm S'\gsim 1$ there is just one value of ${\rm Rm}'$ for marginal instability so that for {\em all}  magnetic 
fields (above a  lower limit) there is one  rotation rate above which the axisymmetric MRI exists.  For a certain 
$\rm S'_{\rm min}$ the ${\rm Rm}'$ takes its overall-minimum ${\rm Rm}'_{\rm min}$. For all Pm between $10^{-4}$ and $10^{3}$ 
the modes with the lowest ${\rm Rm}'$ (which are easiest to excite)  are axisymmetric. 

The dependence of the  curves of marginal instability on the magnetic Prandtl number ${\rm Pm}$  is only weak. For $\rm Pm<1$ 
there is  no visible  influence of ${\rm Pm}$. Axisymmetric global MRI even exists for very small Pm with one and the same 
magnetic Reynolds number (but, of course, the ordinary Reynolds number takes very high values).   

For $\rm Pm>1$ the behavior is different: both the critical ${\rm Rm}'_{\rm min}$ and ${\rm S}'_{\rm min}$ grow with 
$\sqrt{\rm Pm}$. Hence, the scaling for $\rm Pm>1$ switches to the parameters
\beg
{\rm Ha}=\frac{B_z H}{\sqrt{\mu_0\rho \nu \eta}}
\label{HA}
\ende
and
\beg
{\rm Rm^*}= \frac{\Om H^2}{\sqrt{\nu\eta}}
\label{REM}
\ende
(Kitchatinov \&  R\"udiger 2004).
 Both the critical rotation rate and magnetic field now run with $\sqrt{\nu\eta}$ instead with 
$\eta$  as it is true for  $\rm Pm<1$. Both expressions are identical for $\rm Pm=1$. Again, the ratio of 
both quantities is free of the two dissipation parameters.
The ratio of the linear rotation velocity and the Alfv\'en velocity of the vertical field is called the magnetic Mach number
$
\rm Mm= {U_\phi}/{V_{\rm A}}
$
with 
\beg
V_{\rm A}=\frac{B_z}{\sqrt{\mu_0\rho}}.
\label{Va}
\ende
At a distance of 1 AU and for a magnetic field of 1 Gauss the magnetic Mach number is of order 100. As the possible magnetic 
field should be smaller than 1 Gauss  this value of Mm is certainly a {\em minimum}.

The behavior of the nonaxisymmetric modes  is not so well-known. We have learnt from the theory of the azimuthal magnetorotational 
instability (AMRI) that too fast   rotation always destroys the instability.  The AMRI follows from the interplay of differential 
rotation and an toroidal current-free magnetic field. It is basically nonaxisymmetric. For a given Lundquist number S there are 
two critical Reynolds numbers Rm. The instability is supercritical only between the two values. With other words, AMRI  is excited 
by  fast enough rotation but   it is suppressed by too fast rotation. Both the weak-field branch and the strong-field branch have 
the same positive slope in the plane Rm over S. The reason for this expressive  phenomenon is the smoothing influence of differential 
rotation on nonaxisymmetric magnetic perturbations.  

The question arises whether also the non\-axi\-symmetric modes of standard MRI are finally supported by the differential rotation. The 
answer has consequences for i) possible dynamo models but also ii) for the magnetic field amplitudes necessary for the excitation 
of nonaxisymmetric modes. Let us assume for a moment that the weak-field branch of the instability map fulfills the condition
\beg
{\rm Rm} \simeq {\rm Mm_{\rm weak} \, S}
\label{b}
\ende
then 
\beg
\frac{B_z}{\sqrt{\mu_0\rho}}> \frac{U_\phi}{\rm Mm_{\rm weak}} \, \frac{H}{R}
\label{bb}
\ende
for the minimum seed field $B_z$  necessary for the excitation of the $m=1$ mode. $\rm Mm_{\rm weak}$ is the (positive) slope of 
the curve according to the calculations. The result does not depend on the actual value of the magnetic diffusivity. 

The same is true for the strong-field branches shown in Fig. \ref{f1}. One finds the corresponding magnetic Mach number 
$\rm Mm_{strong}$ much smaller than $\rm Mm_{weak}$. For given basic rotation the flow is unstable against nonaxisymmetric 
perturbations if
\beg
{\rm Mm_{strong}<Mm< Mm_{weak}}.
\label{Mm}
\ende

For Kepler disks the relation (\ref{bb}) can be read as
\beg
\beta \lsim {\rm Mm_{\rm weak}}^2,
\label{seeddisk}
\ende
with $\beta=\mu_0 P/B^2_z$ as the plasma $\beta$. Hence, the $\beta$ must be rather small to excite nonaxisymmetric modes. 
The value of 400 used by Fromang et al. (2007) is so high that the corresponding magnetic fields could be much too 
weak to excite nonaxisymmetric instability modes. Kitchatinov \& R\"udiger (2010) for a simplified Kepler disk find the rather 
moderate values $\rm Mm_{strong}\simeq 2$ and $\rm Mm_{weak}\simeq 10$. For such small values the unstable domain of the 
nonaxisymmetric modes is rather restricted. In the present paper we shall find that the numerical values of the critical magnetic 
Mach numbers for the weak-field branch for modes in infinite cylinders are much higher.

%%%%%%%%%%%%%%%%%%%%%%%%%%%%%%%%%%%%%%%%%%%%%%%%%%%%%%%%%%%%%%%%%%%%%%%%%%%%%%%%%%%%%%%
\section{The model}
%%%%%%%%%%%%%%%%%%%%%%%%%%%%%%%%%%%%%%%%%%%%%%%%%%%%%%%%%%%%%%%%%%%%%%%%%%%%%%%%%%%%%%%%
The MRI has been found by  considering the stability problem of Taylor-Couette (TC) flows of magnetized ideal fluids by Velikhov (1959). 
Because of the simplicity of this geometry much work has been done to study the MRI for fluids between cylinders rotating with 
different angular velocities. Theory (R\"udiger \& Zhang 2001; Ji et al. 2001) and experiments (see Stefani et al. 2009) revealed 
the possibilities to realize the MRI even in the laboratory.

In the following, two concentric cylinders (unbounded in $z$) of radii $R_{\rm in}$ and  $R_{\rm out}$ are considered with the 
rotation rates $\Om_{\rm in}$ and $\Om_{\rm out}$. 
The rotation profile between the cylinders  may mimic the Kepler rotation law, i.e. we shall fix  $\Om_{\rm out}=0.35 \Om_{\rm in}$. 
The value ensures that the cylinders with $R_{\rm out} = 2  R_{\rm in}$ rotate like planets.  We call this radial rotation profile 
as quasi-Keplerian profile. 

The boundaries are assumed to be impenetrable, stress-free and perfect conducting, which are valid also for almost all liquid metal 
experiments in the laboratory. The cylinders are infinite in the axial direction. Fricke (1969) and Balbus \& Hawley (1990)  started 
to apply the instability to important astrophysical applications. In order to be more close to  the physics of accretion disks we 
also considered  in the nonlinear simulations a pseudo-vacuum condition for the outer boundary. 

The governing equations are
\begin{eqnarray}
\frac{\partial \vec{u}}{\partial t} + (\vec{u} \nabla)\vec{u} =
-\frac{1}{\rho} \nabla P + \nu \Delta \vec{u} + 
\frac{1}{\mu_0}{\textrm{curl}}\ \vec{B} \times \vec{B},
\label{mhd}
\end{eqnarray}
\begin{eqnarray}
\frac{\partial \vec{B}}{\partial t}= {\textrm{curl}} (\vec{u} \times \vec{B})+ \eta \Delta\vec{B},
\label{mhd1}
\end{eqnarray}
and
\beg
{\textrm{div}}\ \vec{u} = {\textrm{div}}\ \vec{B} = 0,
\label{mhd2}
\ende
where $\vec{u}$ is the velocity, $\vec{B}$ the magnetic field, $P$  the 
pressure, $\nu$ the kinematic viscosity, and $\eta$ the magnetic diffusivity.

The  basic state is $u_R=u_z=B_R=B_\phi=0$, $B_z=B_0=$ const. and
\beg
U_\phi=R\Om=a R+\frac{b}{R},
\label{basic}
\ende
where $a$ and $b$  are constants defined by 
\beg
a=\Om_{\rm{in}}\frac{ \mu_\Omega-{\hat\eta}^2}{1-{\hat\eta}^2}, \q
b=\Om_{\rm{in}} R_{\rm{in}}^2 \frac{1-\mu_\Omega}{1-{\hat\eta}^2},
\label{ab}
\ende
with
\begin{equation}
\hat\eta=\frac{R_{\rm{in}}}{R_{\rm{out}}}, \quad\quad\quad
\mu_\Omega=\frac{\Om_{\rm{out}}}{\Om_{\rm{in}}} .
\label{mu}
\end{equation}

The axial field amplitude is now measured by the  Hartmann number
\beg
{\rm Ha} = \frac{B_0 D}{\sqrt{\mu_0 \rho \nu \eta}}.
\ende
$D=R_{\rm out} - R_{\rm in}$  is used as the unit of length, $\nu/D$ as the unit of velocity. 
The rotation $\Om$ is normalized with the inner rotation rate $\Om_{\rm in}$. The magnetic Reynolds number $\rm Rm$ is defined as 
\beg
{\rm Rm}=\frac{\Om_{\rm in}  D^2}{\eta}.
\label{Rey}
\ende
The Lundquist number S is defined by $\rm S=Ha \cdot \sqrt{\rm Pm}$. Expressed with the 
Alfv\'en frequency $\Omega_{\rm A}$ it is 
\beg
\frac{\Omega_{\rm A}}{\Omega_{\rm in}} = \frac{\rm S}{\rm Rm}.
\label{OmA}
\ende
All the calculations in this paper are done for a model with $\hat\eta=0.5$. This is the only geometry where the scales 
$R_{\rm in}$ and $D$ are equal. As the magnetic Mach number the quantity $\rm Mm=\Omega_{\rm in}/\Omega_{\rm A}$ is used.

%%%%%%%%%%%%%%%%%%%%%%%%%%%%%%%%%%%%%%%%%%%%%%%%%%%%%%%%%%%%%%%%%%%%%%%%%%%%%
\section{Wave number and drift frequencies}
%%%%%%%%%%%%%%%%%%%%%%%%%%%%%%%%%%%%%%%%%%%%%%%%%%%%%%%%%%%%%%%%%%%%%%%%%%%%%%%%%%%%%%%%%%%%
The equations (\ref{mhd} --  \ref{mhd2}) are linearized with respect to the background state (\ref{basic}). The perturbed quantities 
are developed after azimuthal Fourier modes 
\beg 
F = F(R) {\rm e}^{{\rm i}(kz+m\phi+\omega t)}.
\label{Fourier}
\ende
The results are optimized in the wave number $k$. Only the solutions with those  $k$ are of interest for which the Reynolds numbers 
take a minimum. All solutions with another $k$ have higher Reynolds numbers. One can also   show that the solutions with a certain 
positive $k$ are always accompanied by a solution with $-k$ but with  the same Reynolds number and drift frequency (for given $\rm Ha$ 
and $m$). As the pitch angle of the resulting spirals is given by $\partial z/\partial \phi= -m/k$ it is clear that both the solutions 
have opposite pitch angles so that the solution is always a combination of a left screw and a right screw.

\begin{figure}[h]
   \centering
   \includegraphics[width=6.0cm]{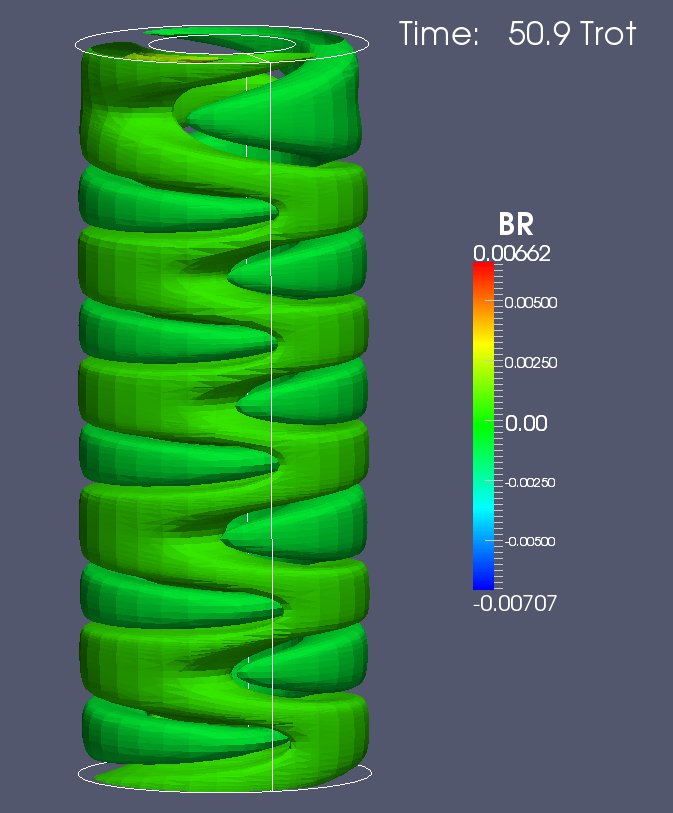}
   \caption{The simultaneous existence of the two modes with $k$ and $-k$ in the nonlinear regime. The whole pattern drifts 
            in the direction of rotation. $\rm Rm=160$, $\rm S=13$,  $\rm Pm=1$. }
   \label{hel}
\end{figure}

The proof is as follows. After elimination of both pressure fluctuations and the fluctuations of the vertical magnetic field, $B'_z$, 
the linearized equations are
\beg
{\partial u_R \over \partial R} + {u_R \over R} + {{\rm i}m \over R} u_\phi + {\rm i}k u_z = 0,
\label{p1}
\ende
\begin{eqnarray}
\lefteqn{{\partial^2 u_\phi \over \partial R^2} + {1\over R} {\partial u_\phi
\over \partial R} - {u_\phi \over R^2} - \left({m^2 \over R^2} + k^2\right)
u_\phi -}\nonumber\\
\lefteqn{-{\rm i} \left(m {\rm Re} {\Om \over \Om_{\rm in}} + \omega\right)
u_\phi + {2{\rm i}m \over R^2} u_R - {\rm Re} {1\over R} 
{\partial \over \partial R}
\left(R^2 {\Om \over \Om_{\rm in}}\right) u_R} \nonumber\\
&& - {m \over k} \left[{1\over R} {\partial^2 u_z \over \partial R^2} + 
{1\over R^2} {\partial u_z \over \partial R} - \left({m^2 \over R^2} + 
k^2\right) {u_z
\over R} - \right. \nonumber\\
&& \left. - {\rm i}\left(m {\rm Re} {\Om \over \Om_{\rm in}} + \omega\right) 
{u_z
\over R}\right]  + {m\over k} {\rm Ha}^2 \left[{1\over R} {\partial B_R \over 
\partial R} +
{B_R \over R^2}\right] +\nonumber\\
&& + {{\rm i}\over k} {\rm Ha}^2 \left({m^2\over R^2} + k^2\right)
B_\phi = 0,
\label{p2}
\end{eqnarray}
\begin{eqnarray}
\lefteqn{{\partial^3 u_z \over \partial R^3} + {1\over R} 
{\partial^2 u_z \over \partial
R^2} - {1\over R^2} {\partial u_z \over \partial R} - \left({m^2\over R^2} +
k^2\right) {\partial u_z \over \partial R} +}\nonumber\\
\lefteqn{+{2m^2 \over R^3} u_z - {\rm i}\left(m
{\rm Re} {\Om\over \Om_ {\rm in}} + \omega\right) {\partial u_z \over
\partial R} -  
 {\rm i}m {\rm Re} {\partial \over \partial R} \left({\Om \over \Om_{\rm
in}}\right) u_z} \nonumber\\ 
&& - {\rm Ha}^2 \left[{\partial^2 B_R \over \partial R^2} + {1\over
R} {\partial B_R \over \partial R} - {B_R \over R^2} - k^2 B_R + \right. \nonumber\\
&& \left. +{{\rm i}m \over R}
{\partial B_\phi \over \partial R} - {{\rm i}m\over R^2} B_\phi \right] -
{\rm i}k\left[{\partial^2 u_R \over \partial R^2} + {1\over R} {\partial u_R
\over \partial R} - {u_R \over R^2} - \right. \nonumber\\
&& \left. - \left(k^2 + {m^2\over R^2}\right)
u_R\right] - k \left(m {\rm Re} {\Om \over \Om_{\rm in}} + \omega\right)
u_R -\nonumber\\
&& - 2 {km \over R^2} u_\phi - 2 {\rm i}k {\rm Re} {\Om \over \Om_{\rm in}}
u_\phi = 0,
\label{p3}
\end{eqnarray}
\begin{eqnarray}
\lefteqn{{\partial^2 B_R \over \partial R^2} + {1\over R} {\partial B_R \over 
\partial R}
- {B_R \over R^2} - \left({m^2 \over R^2} + k^2\right) B_R -}\nonumber\\
&& - {2{\rm i}m\over R^2}
B_\phi - {\rm i} {\rm Pm} \left(m {\rm Re} {\Om \over \Om_{\rm in}}
+\omega\right) B_R + {\rm i}k u_R=0,
\label{p4}
\end{eqnarray}
\begin{eqnarray}
\lefteqn{{\partial^2 B_\phi \over \partial R^2} + {1\over R} 
{\partial B_\phi \over
\partial R} -{B_\phi \over R^2} - \left({m^2 \over R^2} + k^2\right) B_\phi
+}\nonumber\\
&& +{2{\rm i}m \over R^2} B_R -{\rm i} {\rm Pm} \left(m {\rm Re} 
{\Om \over \Om_{\rm in}}
+ \omega\right) B_\phi + {\rm i}k u_\phi+\nonumber\\ 
&& + {\rm Pm} \ {\rm Re} \ R {\partial \Om
/\Om_{\rm in} \over \partial R} B_R = 0
\label{p5}
\end{eqnarray}
(Shalybkov et al. 2002). The eigenfrequency $\omega$ is here normalized with the viscosity frequency  $\nu/R_0^2$.
One finds the system as invariant against the simultaneous transformation $k\to -k$, $u_z\to -u_z$, $B_R\to -B_R$ 
and $B_\phi\to -B_\phi$. Hence, if a solution is known for a certain $k$, then always a   modified solution exists 
for $-k$. The standard MRI in cylindric geometry always produces the same number of left and right 
screws\footnote{The same is true for the instability of toroidal fields between the cylinders without and with 
electric current, see R\"udiger et al. (2011).}. The total helicity (kinetic and magnetic) is thus vanishing (Fig. \ref{hel}).

Conjugating the linearized  equation system (\ref{p1} -- \ref{p5})  leads to the  finding that the system is invariant 
against the simultaneous transformations $m\to -m$, $k\to -k$ and $\Re{(\omega})\to -\Re{(\omega})$. Hence, both the 
pattern drift $\partial\phi/\partial t=-\Re{(\omega)}/m$ of both solutions and also the pitch 
\beg
\frac{\partial z}{\partial \phi}= -\frac{m}{k} 
\label{pitch}
\ende
are equal so that both the solutions are identical. It is thus enough  to consider the solutions for positive $m$.

The unstable Taylor-Couette flow forms axisymmetric or nonaxisymmetric  vortices.  
With our normalization the vertical extent $\delta z$ of a  vortex is 
given by
\beg
{\delta z \over R_{\rm out} - R_{\rm in}} = {\pi \over k}
\sqrt{{\hat \eta \over 1-\hat\eta}},
\label{delz}
\ende
hence for  $\hat\eta=0.5$
\beg
\frac{\delta z }{ R_{\rm out} - R_{\rm in}} = \frac{\pi}{ k}.
\label{delta}
\ende
 For  $k\simeq \pi$
the cells have the same vertical extent as they have in radius and for $k\gg\pi$ the cells are very flat. 
Generally,  the vortices of  the
axisymmetric modes become more and more elongated in the vertical direction, $k\ll\pi$.

The drift velocity ${\cal R} (\omega)$ which always proved
to be negative, i.e. the pattern drifts are in direction of the rotation (eastward). It is
\beg
\dot \phi =-  {{\cal R}(\omega) \Om_{\rm in} \over m },
\label{dotfi}
\ende
so that  the eastern drift period in units of the rotation period is
$m/{\cal R}(\omega)$. Note that we are working in the resting laboratory system.

%%%%%%%%%%%%%%%%%%%%%%%%%%%%%%%%%%%%%%%%%%%%%%%%%%%%%%%%%%%%%%%%%%%%%%%%%%%%%%%
\section{Solutions for Pm=1}
\begin{figure}[h]
   \centering
   \includegraphics[width=9.0cm,height=8cm]{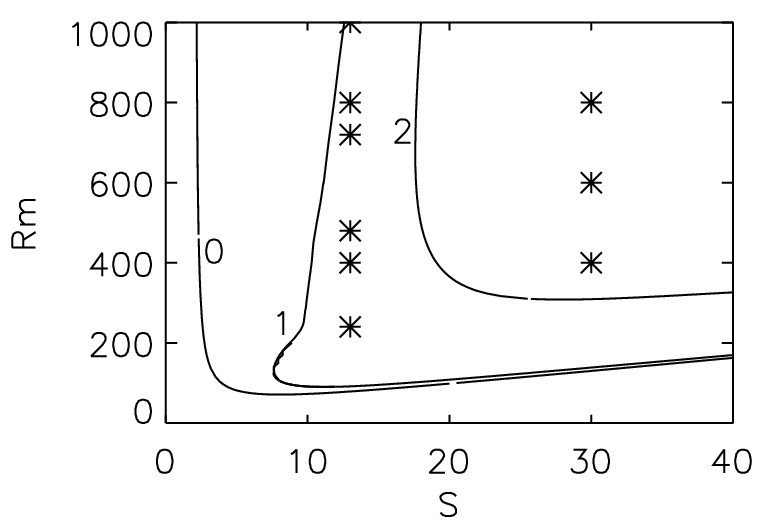}
   \caption{The instability map of the modes with $m=0$, $m=1$ and $m=2$ in quasi-Kepler flows ($\mu_\Omega=0.35$). For 
            given ${\rm Rm}$ the  flow in unstable for Lundquist numbers $\rm S$ between the values at the weak-field 
            branch and the strong-field branch. Star symbols mark the series of nonlinear simulations discussed below. $\rm Pm=1$. }
   \label{f1}
\end{figure}

%%%%%%%%%%%%%%%%%%%%%%%%%%%%%%%%%%%%%%%%%%%%%%%%%%%%%%%%%%%%%%%%%%%%%%%%%%%%%%
We start with the standard case $\rm Pm=1$ where no difference exist between the both Reynolds numbers 
and also no difference exists  between the Hartmann number and the Lundquist number. Most numerical simulations concern 
  this case. The critical Reynolds numbers for excitation of the modes  with $m=0$, $m=1$ and $m=2$ in 
quasi-Kepler flows are given in 
Fig. \ref{f1}. 

For the  given Lundquist numbers  the Reynolds numbers are minimized by variation of the axial wave number $k$. Note 
the existence of an absolute minimum ${\rm Re}_{\rm MIN}$ of the Reynolds number. It is smaller for the axisymmetric 
mode than for the nonaxisymmetric modes. The nonaxisymmetric modes need faster rotation for their excitation. There 
is a basic difference, however, between the axisymmetric and the nonaxisymmetric modes. For $m=0$ and $\rm S\simeq 1$  
exists always {\em one} critical Reynolds number above which the MRI is excited for {\em all} larger Rm. The absolute minimum 
value for S is of order unity. On the other hand, the critical  Lundquist values for the nonaxisymmetric modes with 
$m>0$ behave completely different. For $\rm S> S_{\rm MIN} $ where $\rm S_{\rm MIN}\simeq 1$ is the smallest possible 
Lundquist number there are always {\em two} critical Reynolds numbers between them the nonaxisymmetric MRI modes can only 
exist (Fig. \ref{f1}). Hence, the nonaxisymmetric mode cannot survive if the rotation is too fast. The 
differential rotation excites the MRI but -- if too strong -- it  suppresses its nonaxisymmetric parts. 

The rotational quenching of the nonaxisymmetric parts of the MRI should have serious consequences for the dynamo problem.
After the theorem by Cowling all dynamo models need nonaxisymmetric parts of the magnetic field. 

The $m=1$ mode in Fig. \ref{f1} needs seed fields which are the higher the faster the rotation is. As an estimation one finds
\beg
\Omega_{\rm A} \simeq 3\cdot 10^{-3}\ \Omega_{\rm in} 
\label{seed}
\ende
($\rm Mm_{\rm weak}\simeq 320$). If we take the linear velocity of the Kepler disk at 1 AU as more than 30 km/s (solar system) 
and the density at this place as 10$^{-10}$ g/cm$^3$, then the relation (\ref{seed}) means
\beg
{B_0} > 1\ \  {\rm Gauss},
\label{seed_value}
\ende
which is a very high value at the distance. The value (\ref{seed_value}) is needed to start any MRI-dynamo. It exceeds the above 
mentioned minimum value of 0.1 Gauss for the excitation of axisymmetric modes by one order of magnitude. The rather high 
value of this field strength suggests that indeed the weak-field branch of the MRI bifurcation map is the branch of 
astrophysical relevance rather than the strong-field branch. For the latter one finds $\Omega_{\rm A} \simeq 0.3\ \Omega_{\rm in}$ 
representing  magnetic fields  which are by a factor of hundred (!) stronger than the fields at the weak-field branch. It is 
hard to imagine that such fields are available for the formation of protoplanetary disks.

\begin{figure}[h]
   \centering
  \mbox{
   \includegraphics[height=4.5cm,width=4.5cm]{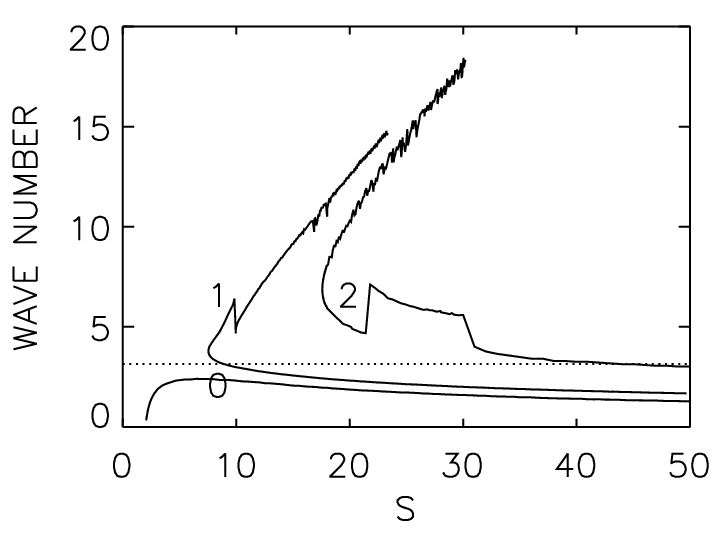}
   \includegraphics[height=4.5cm,width=4.5cm]{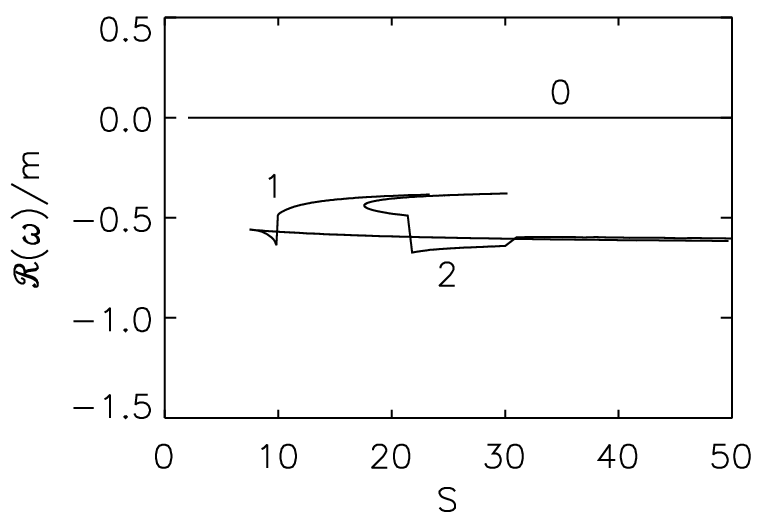}}
   \caption{The same as in Fig. \ref{f1} but for the normalized wave number (left) and the drift frequency (right).}
   \label{f2}
\end{figure}

The results for the wave numbers (for which the Reynolds numbers for fixed S are minimum) are plotted in Fig. \ref{f2} (left). 
The dotted line gives $k=\pi$ for which the cells after (\ref{delta}) are nearly spherical. For $k<\pi$ the cells tend to be 
prolate while they are oblate for $k>\pi$. This is, of course, only true if the radial cell size is 
of the order of the distance between the cylinders which must be checked separately. 

\begin{figure}[h]
   \centering
   \includegraphics[width=8.3cm]{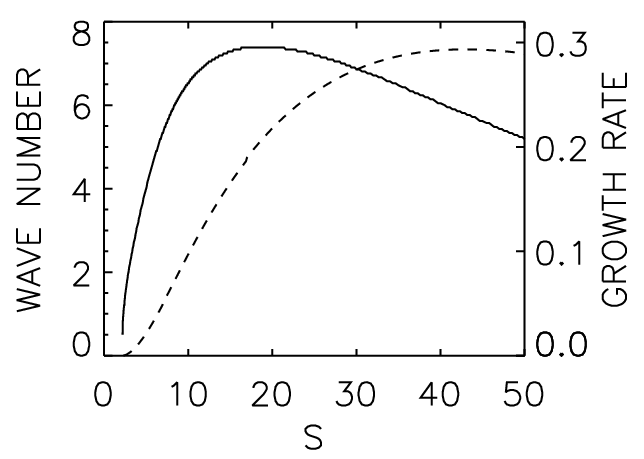}
   \caption{The dependence on the magnetic field of the wave number of the axisymmetric mode  and the growth rate (dashed) 
            for given Reynolds number $\rm Rm=500$. $m=0$, $\rm Pm=1$. }
   \label{wave}
\end{figure}

There are surprising differences in   Fig. \ref{f2} (left) for axisymmetric and nonaxisymmetric modes. All the axisymmetric 
rolls are prolate as their axial size exceeds the radial scale. This is true for all values of the magnetic field. The result 
is not surprising as magnetic fields always increase the correlation length along the field, hence for strong fields we have 
$k\propto 1/{\rm S}$. That for $m=0$ the wave numbers $k$ for given $\rm Rm$ are small at both  the weak-field (left)  branch 
and the strong-field (right) branch of the marginal instability curve does {\em not} mean that they are small also between. 
Kitchatinov \& R\"udiger (2004) find with a  local approximation close to  the left branch the relation $k\simeq V_{\rm A}/\eta$ 
so that the wave number runs with $\rm S$. One must thus expect that between the left branch  and the right branch there is a 
maximum of $k$ so that there also the axisymmetric modes have short axial scales -- as we shall demonstrate by Fig. \ref{f71} 
(top, right). Figure \ref{wave} gives for $\rm Rm=500$ the mentioned maximum of the axial wave number together with the growth 
rate. One finds that the axisymmetric channel mode forms thin rolls only close to the left branch but the rolls are thick for 
stronger  fields, i.e  values of $\rm Mm\gsim 1$.

For $m=1$ only the cells with large $\rm S$ are almost rolls. The wave numbers for both  branches are rather different. They 
are  large for the weak-field  branch and they are  small for the strong-field  branch. The cells for the marginal modes 
along the weak-field  branch  are thus rather flat. This is only true, however, if the radial eigenfunctions do smoothly 
cover the distance between the cylinders.

The drift of the nonaxisymmetric modes is eastward  typically 50\% of the rotation frequency of the inner cylinder, i.e. the 
MRI pattern drift is westward with respect to the rotating system. As the outer cylinder only rotates with 35\% of the rotation 
frequency, the field pattern rotates by 42\% faster than   the outer cylinder. 
The corotation radius of the MRI pattern is located between both the cylinders close to  the middle  between the cylinders. 
The same is true, at least qualitatively, for the pattern with $m=2$. A test calculation with $\hat\eta=0.8$ leads to a pattern 
drift $-\Re{(\omega)}/m \simeq 0.85$ which also yields  a corotation radius just in the middle of the two cylinders.

%%%%%%%%%%%%%%%%%%%%%%%%%%%%%%%%%%%%%%%%%%%%%%%%%%%%%%%%%%%%%%%%%%%%%%%%%%%%%%%
\section{Solutions for Pm=0.01}
\begin{figure}[h]
   \centering
   \includegraphics[width=9.0cm]{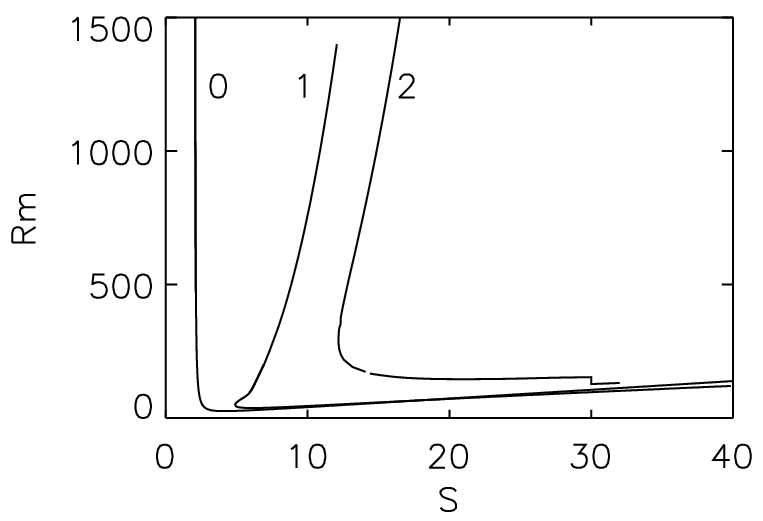}
   \caption{The same as in Fig. \ref{f1} but for $\rm Pm=0.01$. At the strong-field branch for $\rm S>20$ the nonaxisymmetric 
            mode becomes the mode with the lowest magnetic Reynolds number. }
   \label{f3}
\end{figure}

\begin{figure}[h]
   \centering
  \mbox{
   \includegraphics[height=4.5cm,width=4.5cm]{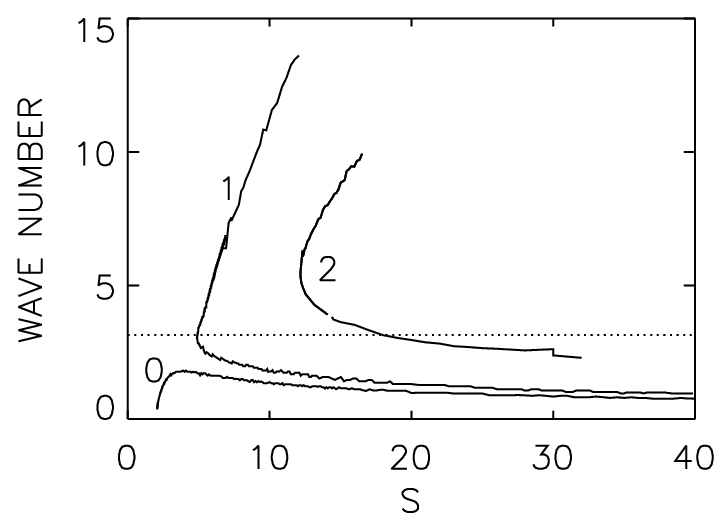}
   \includegraphics[height=4.5cm,width=4.5cm]{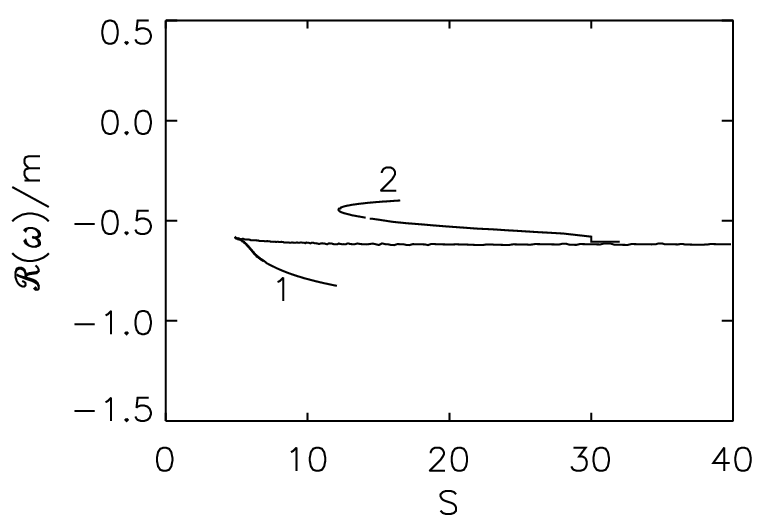}}
   \caption{The normalized wave number (left) and the drift frequency (right) for $\rm Pm=0.01$.}
   \label{f4}
\end{figure}

%%%%%%%%%%%%%%%%%%%%%%%%%%%%%%%%%%%%%%%%%%%%%%%%%%%%%%%%%%%%%%%%%%%%%%%%%%%%%%
The results for smaller magnetic Prandtl numbers ($\rm Pm=0.01$) are given in the plots \ref{f3} and \ref{f4}. In the weak-field 
limit the differences to $\rm Pm=1$ are rather small. The rotational quenching of the nonaxisymmetric modes seems to be slightly 
weaker than for larger Pm ($\rm Mm_{\rm weak}\simeq 400$). 

Also the behavior of the wave number and the azimuthal drift is rather obvious. The cells in the weak-field limit are  flat while 
they are highly aligned along the rotation axis in the strong-field limit (Fig. \ref{f4}, left). The axisymmetric cells are longer 
than the nonaxisymmetric cells.

More striking is the phenomenon for strong fields that the marginal-stability curves for $m=0$  and $m=1$ are crossing 
for $\rm S\simeq 18$. The mode with the global-minimum Reynolds number is always axisymmetric but for stronger magnetic 
fields the critical Reynolds numbers for $m=1$ are smaller than those for $m=0$. For such fields and for increasing Reynolds 
numbers MRI sets in as a nonaxisymmetric flow pattern. The nonaxisymmetric structure is lost, however, for too fast rotation when 
the magnetic Reynolds number reaches the upper value of the marginal instability of the $m=1$ mode and the solution becomes axisymmetric 
again. We have found this sort of mode crossing both for different geometries (Kitchatinov \& R\"udiger 1997) and also with 
different codes (Shalybkov et al. 2002). So far this phenomenon is only known for MHD flows between conducting cylinders.

%%%%%%%%%%%%%%%%%%%%%%%%%%%%%%%%%%%%%%%%%%%%%%%%%%%%%%%%%%%%%%%%%%%%%%%%%%%%%%%%
\section{The magnetic Prandtl number dependence}
\begin{figure}[htb]
   \centering
   \mbox{
   \includegraphics[height=4.5cm,width=4.5cm]{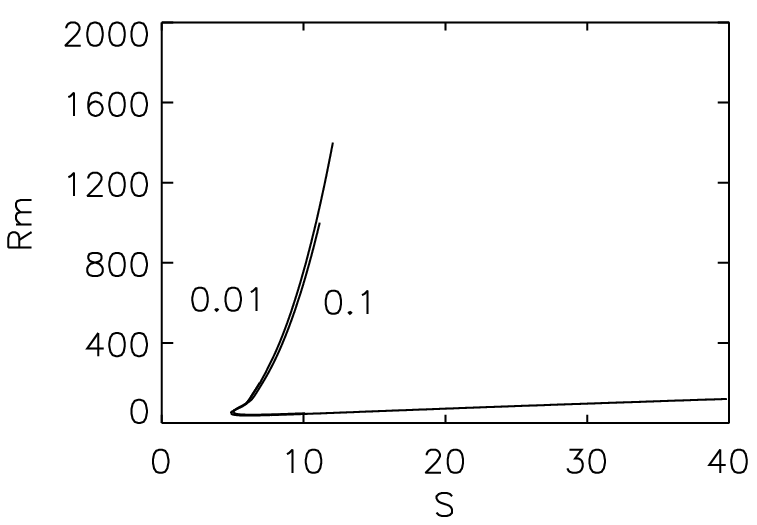}
   \includegraphics[height=4.5cm,width=4.5cm]{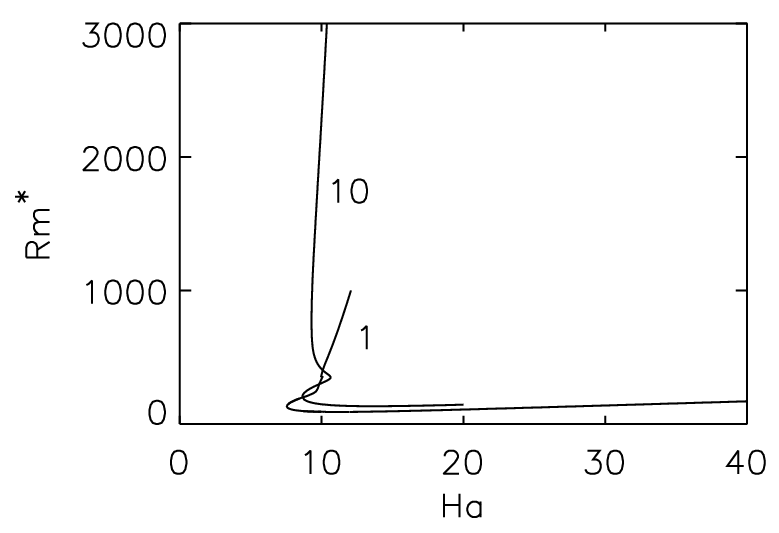}}
   \caption{The critical Reynolds numbers for the nonaxisymmetric modes with $m=1$ for small (left) and 
            large (right) magnetic Prandtl numbers. }
   \label{f5}
\end{figure}

%%%%%%%%%%%%%%%%%%%%%%%%%%%%%%%%%%%%%%%%%%%%%%%%%%%%%%%%%%%%%%%%%%%%%%%%%%%%%%
There are simple scaling laws for the  marginal instability. They appear if also the instability domains for
large magnetic Prandtl numbers  are computed. The results for $\rm Pm=10$  are plotted with the variable $\rm Rm^*$ 
and Ha (Fig. \ref{f5}, right). Both the Figs. \ref{f5} demonstrate the small differences of the curves 
for small Pm if scaled with Rm and S, while the same is true for the curves for large Pm if scaled 
with $\rm Rm^*$ and Ha. In both cases the ratio of the characteristic numbers is the magnetic Mach number which is free of 
any diffusivity values. We find a somewhat weaker rotational quenching 
\beg
\Omega_{\rm A} \simeq 3\cdot 10^{-4}\ \Omega_{\rm in} 
\label{seed1}
\ende
of the nonaxisymmetric modes for  $\rm Pm>1$ compared with the relation (\ref{seed})  which holds for $\rm Pm\leq 1$. 
It is the  general behavior of the weak-field edges of the instability domains for $\rm Pm\lsim 1$ to show a positive 
slope. However, after inspection of Fig. \ref{f5} (right)  for $\rm Pm=10$ the slope is very large and it is unclear 
whether it is still positive. In any case the behavior of the weak-field limit for large Prandtl numbers slightly 
differs from  that of the curves for smaller Pm. The differences provided by the strong-field branches are very small 
for the two regimes.

%%%%%%%%%%%%%%%%%%%%%%%%%%%%%%%%%%%%%%%%%%%%%%%%%%%%%%%%%%%%%%%%%%%%%%%%%%%%%%%%
\section{Nonlinear simulations}
%%%%%%%%%%%%%%%%%%%%%%%%%%%%%%%%%%%%%%%%%%%%%%%%%%%%%%%%%%%%%%%%%%%%%%%%%%%%%%
With the 3D spectral MHD code for cylindric geometry described by Gellert et al. (2007) also nonlinear simulations of global 
MRI are possible. The code works with $M$ Fourier modes in the azimuthal direction, a typical value used in the calculations 
is $M=16$ for the rather narrow spectrum of excited modes.

%%%%%%%%%%%%%%%%%%%%%%%%%%%%%%%%%%%%%%%%%%%%%
\subsection{The instability pattern}
%%%%%%%%%%%%%%%%%%%%%%%%%%%%%%%%%%%%%%%%%%%%%%%%%%%%
The simulations concerning the instability pattern are related  to the map of marginal instability for $\rm Pm=1$ 
(Fig. \ref{f1}) for a fixed Lundquist number $\rm S=13$. Three examples are given. The first one works with relatively 
slow rotation  while the second container  rotates  faster  or much faster. The marginal instability appears for a 
minimum magnetic Mach number of about 6. The first model   lies below  the instability domain of $m=1$, the second 
one inside the $m=1$ domain  and the third model lies outside. The magnetic Mach numbers for the considered cases  
are $\rm Mm=7$, $\rm Mm=46$ and $\rm Mm=96$. 

\begin{figure*}[t]
   \centering
   \vbox
   {\mbox
   { \includegraphics[width=5.0cm]{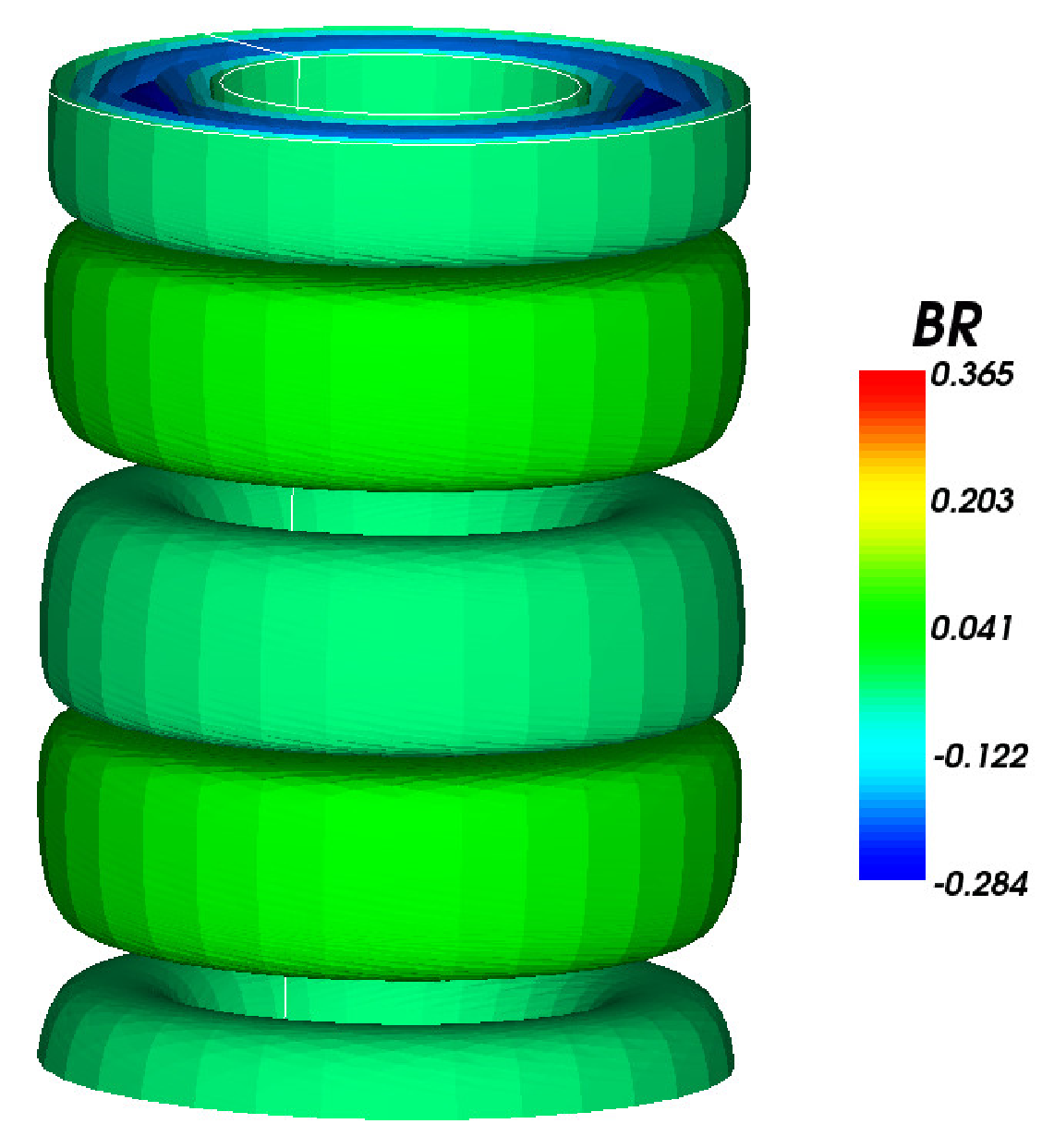}
   \hspace*{4mm}
   \includegraphics[width=5.0cm]{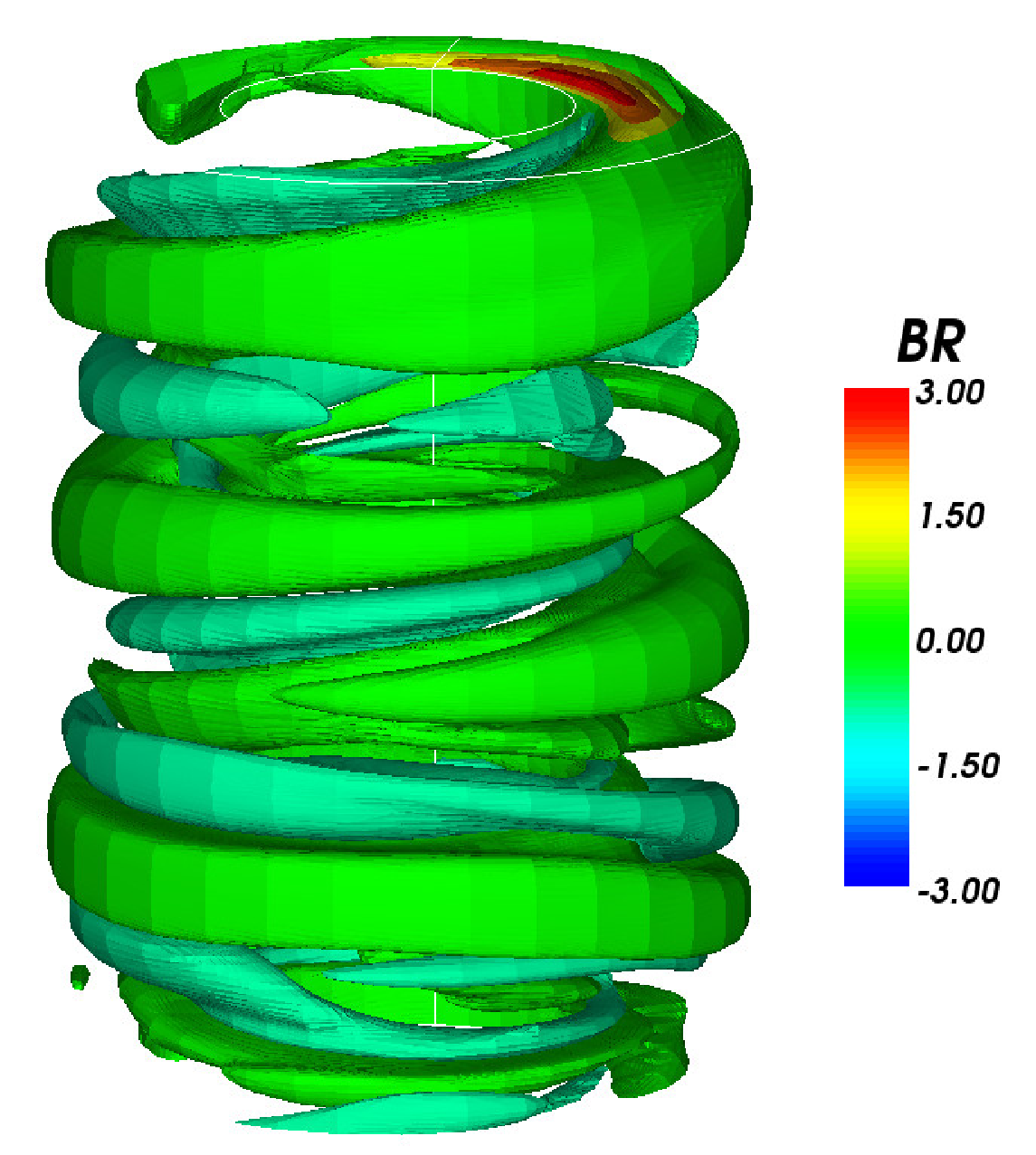}
   \hspace*{4mm}
   \includegraphics[width=5.0cm]{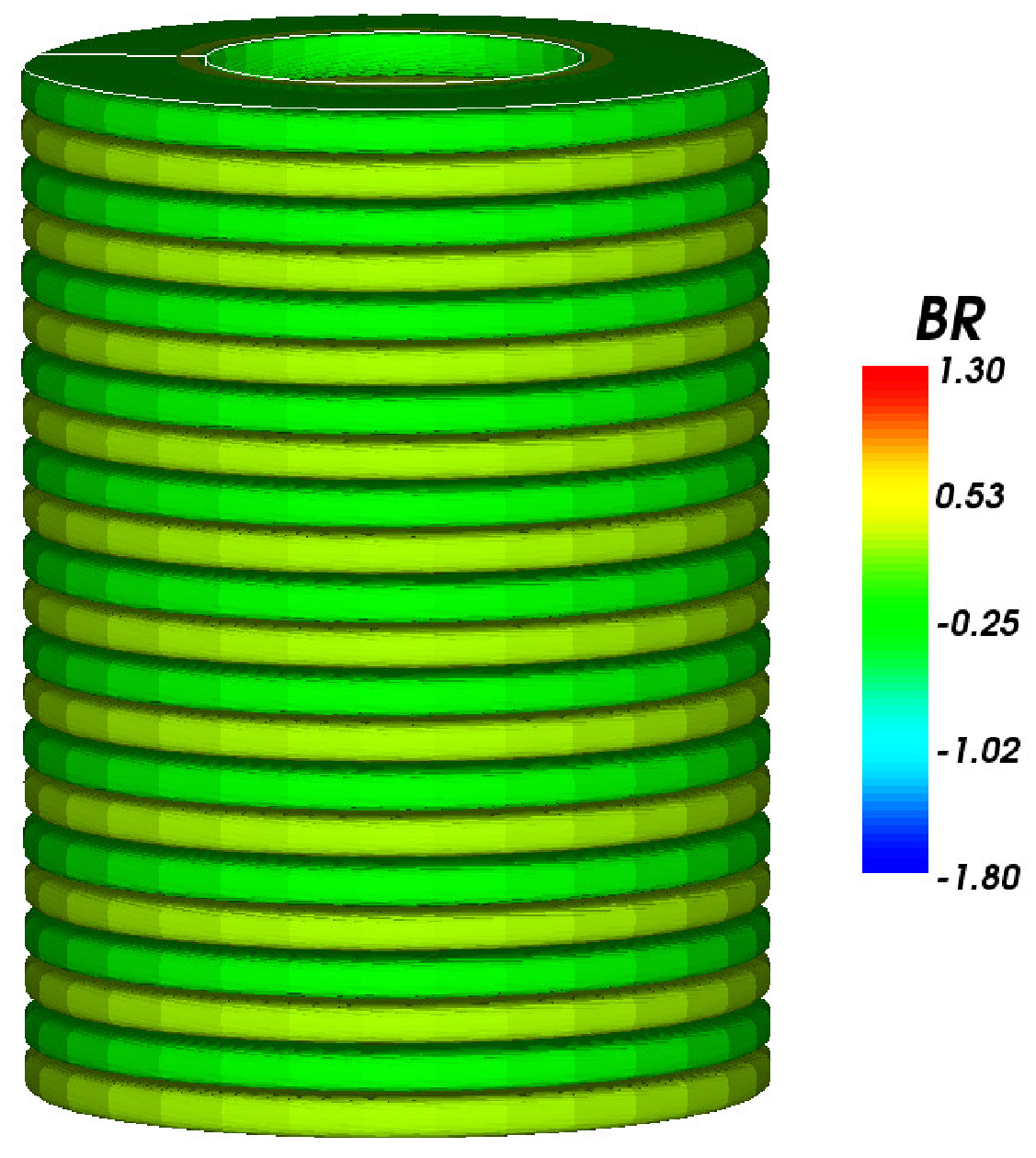}}}

   \vspace*{2mm}

   {\mbox
   {\includegraphics[width=5.0cm,height=4.2cm]{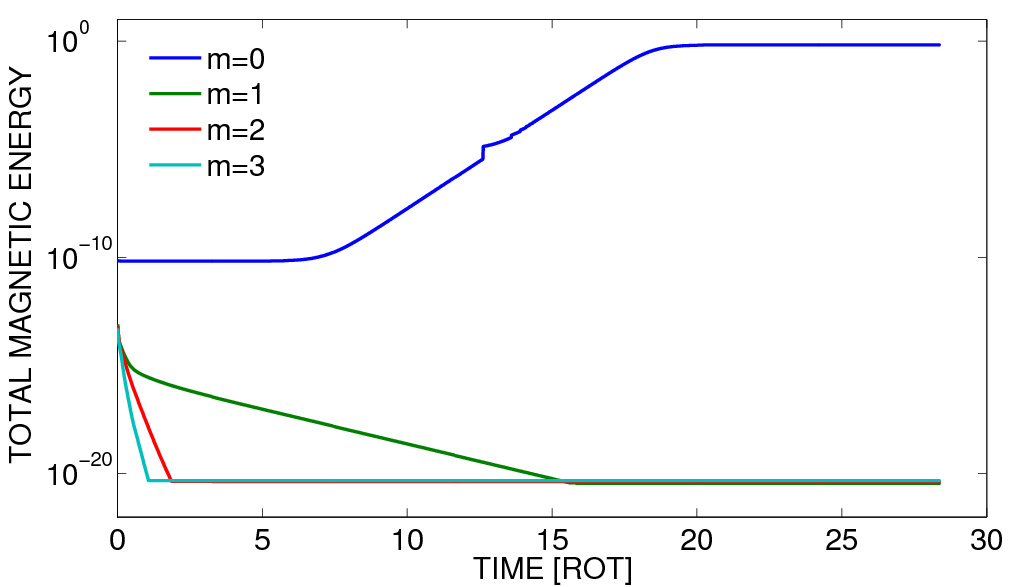}
   \hspace*{4mm}
   \includegraphics[width=5.0cm,height=4.2cm]{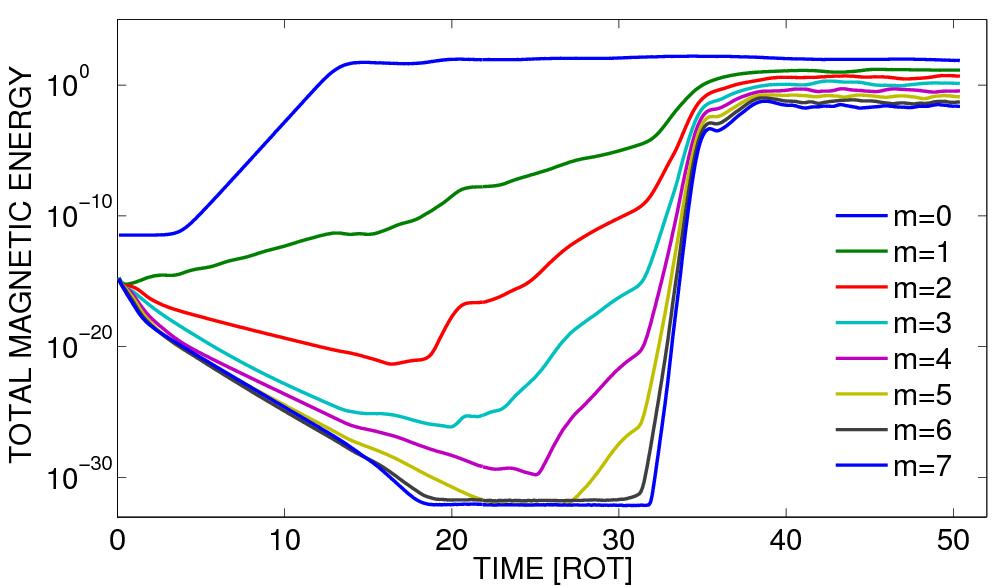}
   \hspace*{4mm}
   \includegraphics[width=5.0cm,height=4.2cm]{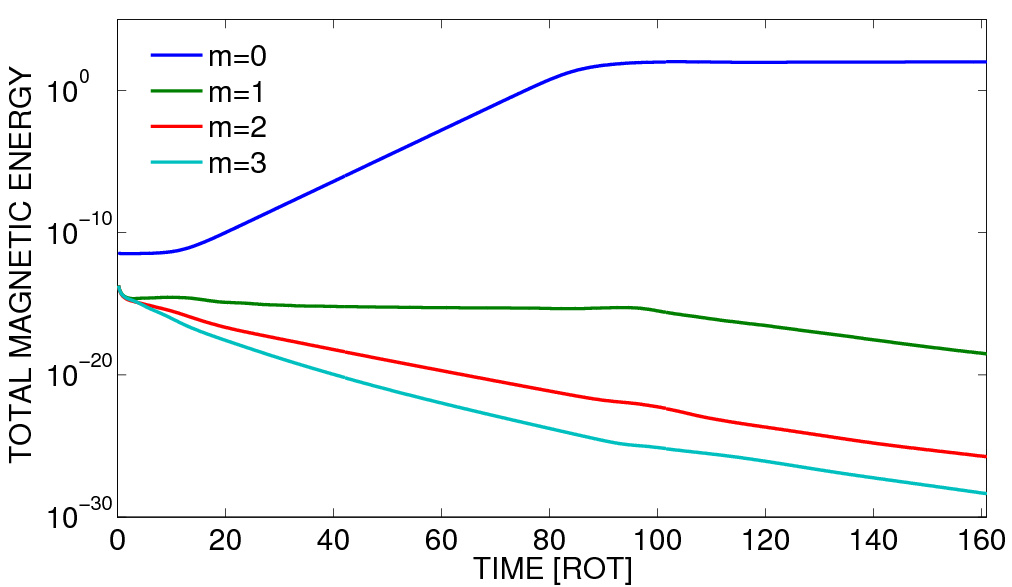}}}
   \caption{Top: the radial component of the magnetic field for $\rm S=13$ and for slow rotation ($\rm Rm= 88$, left), 
            medium rotation  ($\rm Rm=600$, middle) and fast rotation  ($\rm Rm=1250$, right). For slow and for fast 
            rotation the nonlinear instability pattern is purely axisymmetric and for medium rotation it is nonaxisymmetric. 
            Bottom: the same for the energies in the Fourier modes of the magnetic field. $\rm Pm=1$.  }
   \label{f71}
\end{figure*}

The nonlinear calculations are  important as our curves for marginal instability of nonaxisymmetric modes only 
concern the stability behavior of the Kepler flow under the presence of the axial field. It is also possible 
that the rolls of the $m=0$ solution become unstable against disturbances with $m>0$ leading to  secondary 
instabilities with nonaxisymmetric  patterns.

Figure \ref{f71} shows the results. For given magnetic field amplitude ($\rm S=13$)  simulations with $\rm Rm=88$,  
$\rm Rm=600$ and one with $\rm Rm=1250$ are  presented. Only the value  $\rm Rm=600$  lies  in the instability 
map (Fig. \ref{f1})  between the lower and the upper limit for nonaxisymmetric instability and only in this case  
a nonaxisymmetric (drifting) magnetic  pattern results (\ref{f71}, middle). From the beginning on, the mode with 
$m=0$ grows fastest and also the mode with $m=1$ grows continuously. The other modes come much later so that only 
after 30 orbits the complete pattern occurs.  

The opposite is true for faster rotation (Fig. \ref{f71},  right).  Here only the mode with $m=0$  grows 
while all the nonaxisymmetric modes decay (Fig. \ref{f71}, bottom right). The resulting magnetic pattern is 
axisymmetric despite of the high value of the Reynolds number. It is thus shown that indeed the linear  
approximation provides  the real instability behavior. For the considered rotation rates a nonaxisymmetric 
instability of the magnetic $m=0$ pattern does not  appear in the numerical simulations.  

Note that -- as predicted by the linear results for the axial wave numbers (Fig. \ref{f2}, left) -- the axisymmetric 
cells for Reynolds numbers close to the instability limit are really prolate (Fig. \ref{f71}, left). They do become 
oblate for the fast-rotation case (Fig. \ref{f71}, right) despite of the action of  the Taylor-Proudman theorem.
The nonaxisymmetric modes  form a  spectrum of modes (see Fig. \ref{f71},  middle). Only models of this kind are used 
in the next Section to compute  the outward angular momentum transport of the MRI. Only for such models  the equations 
form a nonlinear mixture of many azimuthal modes close to the transition of turbulence. Obviously, too fast rotation 
destroys the resulting mixture (Fig. \ref{f71}, bottom right).

The amplitudes of the MRI-induced magnetic fluctuations are also of importance. They easily exceed the strength of the 
axial background field but only in the domain where the nonaxisymmetric modes are excited (see Fig. \ref{f71}). On the 
other hand, the amplitudes of the toroidal field components grow with growing Reynolds number. For $\rm Re= 600$ the 
maximum $B_\phi$ exceeds the axial background field by one order of magnitude.

%%%%%%%%%%%%%%%%%%%%%%%%%%%%%%%%%%%%%%%%%%
\subsection{The angular momentum transport}
\begin{figure}
   \centering
   \includegraphics[width=8.0cm, height=7cm]{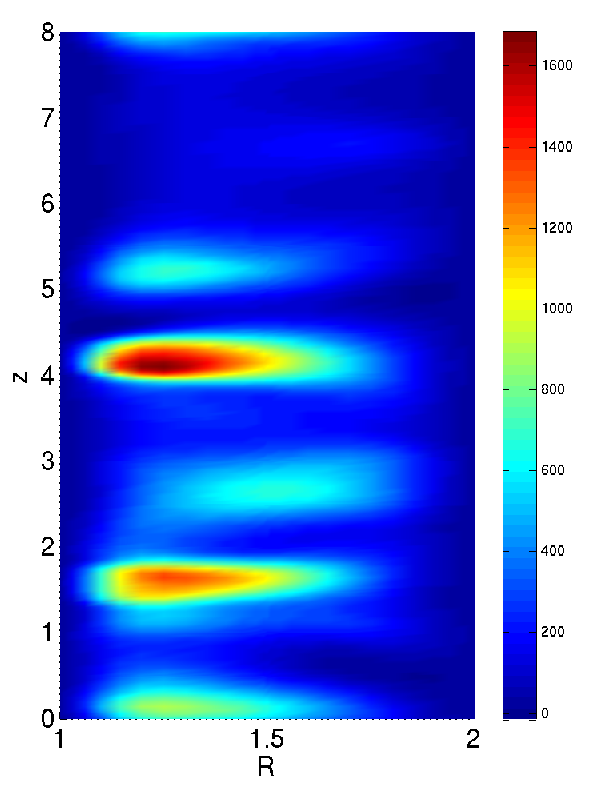}
   \caption{The angular momentum transport parameter $\alpha$ if averaged only over the azimuth is everywhere positive 
            and only slightly reflecting a cell structure. $\rm Rm=400$, $\rm S=30$, $\rm Pm=1$.}
   \label{alphar}
\end{figure}

%%%%%%%%%%%%%%%%%%%%%%%%%%%%%%%%%%%%%%%%%%%
The total angular momentum transport in radial direction is
\begin{equation}
T_{R\phi}= \langle u_R' u_\phi'\rangle - \frac{1}{\mu_0\rho} \langle B_R' B_\phi'\rangle
\label{Trfi}
\end{equation}
which in the usual manner can be written as
\begin{equation}
T_{R\phi}=-\nu_{\rm T} R \frac{d\Omega}{dR}  =\alpha_{\rm SS} \Om^2 D^2
\label{Trfi1}
\end{equation}
with $D$ as the gap width of the container. Hence, the MRI-$\alpha$ as the normalized angular momentum transport can 
be computed with the definition  
\begin{equation}
\alpha_{\rm SS}= \frac{T_{R\phi}}{\Om^2 D^2}.
\label{alfs}
\end{equation}
Note that this definition differs from the usual one unless  $D\simeq H$, which values are indeed of the same order  
for thick disks. In the first step we computed the quantity (\ref{alfs}) by averaging only over the azimuth 
(Fig. \ref{alphar}). One learns that the angular momentum transport is positive everywhere with a weak indication of 
the cell structure. There are no areas in the computational domain with negative angular momentum transport. Generally, 
after our experiences the Maxwell part in the relation (\ref{Trfi}) dominates the Reynolds term.

In order to overcome the axial inhomogeneity we continue to average over the $z$ coordinate so that the resulting 
expression remains only a function of $R$. The resulting profile shows a characteristic maximum as it must vanish 
at the boundaries because of the boundary conditions. The question is how this maximum is related to the mean 
pressure in the computational domain. The pressure in the container vanishes at the inner cylinder and monotonously 
grows towards the outer cylinder. The radially averaged pressure for Kepler rotation  between the bounding cylinders 
is of order  $\rho \Omega^2 D^2$ so that indeed the relation (\ref{Trfi1}) can be read as
\begin{equation}
\rho T_{R\phi}=\alpha_{\rm SS} \ p.
\label{Trfi2}
\end{equation}
This is the standard relation of the accretion theory which -- as we now know --  only holds in the  model after 
averaging over the entire  cylinder.

\begin{figure}[htb]
   \centering
   \vbox{  \includegraphics[width=8.1cm,height=6cm]{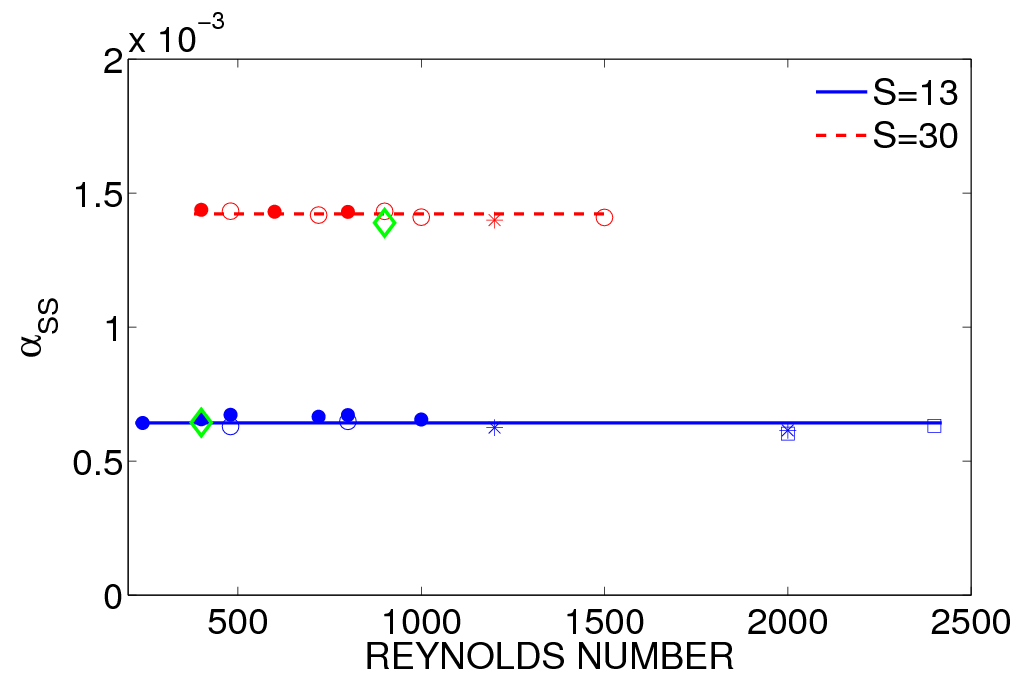}
   \includegraphics[width=7.8cm,height=5.5cm]{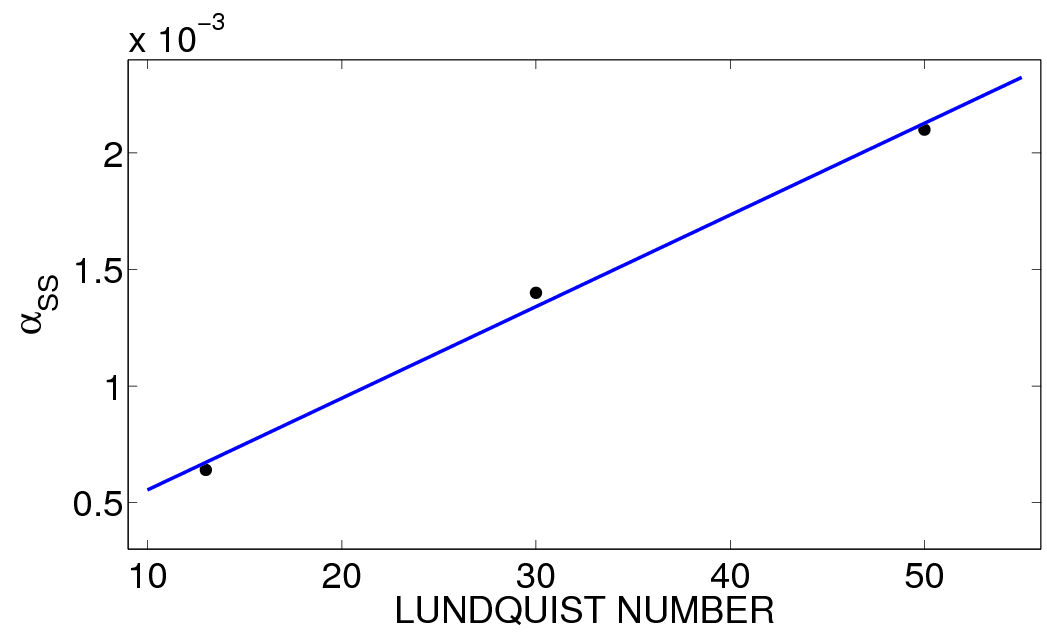}}
   \caption{The normalized angular momentum transport $\alpha_{\rm SS}$ (averaged over the entire cylinder) in its 
            dependence on the Reynolds number (top) and the Lundquist number S (bottom). Influences of both   the 
            viscosity and the basic rotation on the angular momentum transport parameter $\alpha_{\rm SS}$ do not 
            appear.  Dots: $\rm Pm=1.0$, circles: $\rm Pm=0.5$, stars: $\rm Pm=0.2$, square: $\rm Pm=0.1$. Green 
            diamonds: $\rm Pm=1.0$ and pseudo-vacuum as outer boundary condition.}
   \label{alpha_f}
\end{figure}

In the next step, therefore, the averaging procedure concerns the full container. The following results for the  
$\alpha_{\rm SS}$ parameter do only concern to this model.  They are given in Fig.~\ref{alpha_f}, and they can be  
represented  by the {\em linear} relation
\begin{equation}
\alpha_{\rm SS}= 4.8 \cdot 10^{-5}\ {\rm S}, 
\label{alfs1}
\end{equation}
valid for all the considered Reynolds numbers and magnetic Prandtl numbers (${\rm Pm}\leq 1$) given in Fig.~\ref{f1} 
with star symbols. There is no dependence of the $\alpha_{\rm SS}$ on the rotation rate except the chosen value of 
the magnetic Reynolds number lies too close to the boundaries of the instability map  (Fig. \ref{f1}). For two 
examples for $\rm Pm=1$ (green diamonds in Fig. \ref{alpha_f}) the outer boundary condition has been changed from 
perfect conductor to pseudo-vacuum. The numbers do not show a remarkable influence of the boundary conditions on 
the resulting values of $\alpha_{\rm SS}$.

From Eq.  (\ref{alfs1}) one finds with the numbers (3) that
\begin{equation}
\alpha_{\rm SS}\simeq 0.005 \frac{B}{1 {\rm Gauss}}\ \frac{D}{\rm 1 AU}. 
\label{alfs3}
\end{equation}
Hence, the numerical value of the MRI-$\alpha$ linearly depends on the amplitude of the magnetic field and/or  the 
size of the disk or the torus. According to the definition (\ref{Trfi1}) one can also understand our   
$\alpha_{\rm SS}$ as a realization of the $\beta$ viscosity in the sense of Duschl et al. (2000) and 
Hur\'{e} et al. 2001). It is obvious that the magnetic field amplitude must not much smaller than about one Gauss 
in order to get $\alpha_{\rm SS}$ values of order $0.01$ or even larger.

The angular momentum transport shown  in Fig. \ref{alpha_f} is only due to the nonaxisymmetric modes with $m>0$. Only 
these modes have been defined  as the `fluctuations' in the definitions of the random functions $\vec{u}$  and 
$\vec{B}$ and the well-defined averaging procedure is considered as the integration over $\phi$. We have shown in 
Fig. \ref{f71} that only these modes in our simulations are close to develop turbulence. 

\begin{figure}[htb]
   \centering
   \includegraphics[width=9.0cm,height=8.2cm]{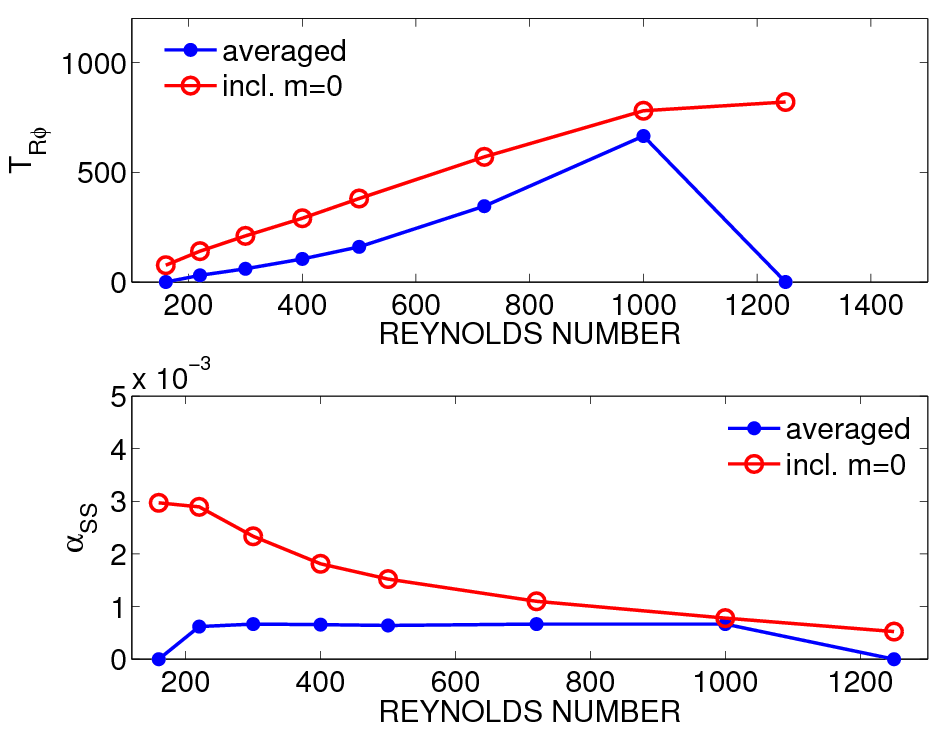}
   \caption{The normalized angular momentum transport $\alpha_{\rm SS}$ (averaged over the entire cylinder) in 
            its dependence on the Reynolds  number. Open circles represent the values if the contribution of the 
            channel mode ($m=0$) is added. $\rm S=13$, $\rm Pm=1$.}
   \label{alpha13}
\end{figure}

\begin{figure}[htb]
   \centering
   \includegraphics[width=9.0cm,height=8.2cm]{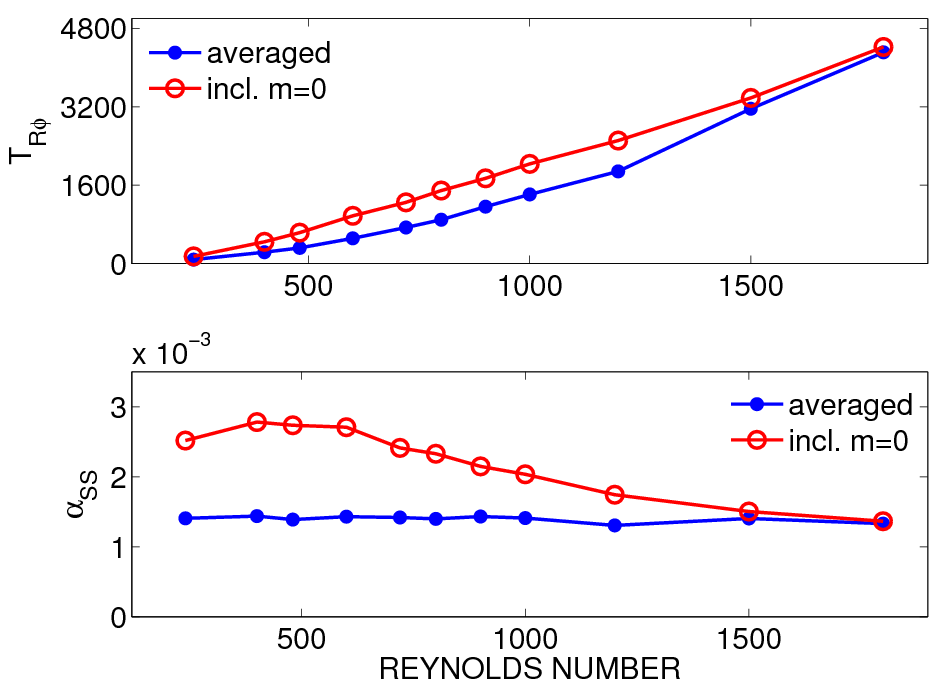}
   \caption{The same as in Fig. \ref{alpha13} but for  $\rm S=30$, $\rm Pm=1$.}
   \label{alpha30}
\end{figure}

The question remains whether also the axisymmetric modes with $m=0$ contribute to the angular momentum transport. They 
can be defined as fluctuations only by their small-scale  spatial vertical variations characterized by the vertical 
wave number $k$. Then the basic rotation can only be  defined by $\langle u_\phi\rangle$ after  averaging over $z$. 
The fluctuations of the flow and the field defined in this way do indeed transport angular momentum. If also the 
axisymmetric modes are used for the calculation of the angular momentum then the resulting values always lie above the 
curves for the nonaxisymmetric modes (Figs. \ref{alpha13} and \ref{alpha30}). For higher Reynolds numbers the extra value 
by the axisymmetric modes diminishes so that the lines in the Figs. \ref{alpha13} and \ref{alpha30} are approaching. In 
this picture the relation (\ref{alfs1})  remains true  if the Reynolds numbers are large enough compared to the 
Lundquist number S, i.e. in the regime of high magnetic Mach number. With such large values of the magnetic Mach number as
discussed above the contributions of the axisymmetric modes to the viscosity-$\alpha$ indeed remain small.

It is shown in Fig. \ref{alpha13} that the angular momentum transport by the nonaxisymmetric modes stops for $\rm Rm \simeq 1200$. 
For larger Reynolds numbers the axisymmetric rolls shown in Fig. \ref{f71} (top, right) do alone produce the viscosity-$\alpha$ 
with almost the same value of about $10^{-3}$. The given Reynolds numbers reach the numerical limits of the code so that the angular 
momentum transport by the $m=0$ mode for very large $\rm Rm$ remains an open question as well as its stability towards larger $\rm Rm$.

Our results do  {\em not} confirm the finding of Lesur \& Longaretti (2007)  and Longaretti \& Lesur (2010) who report strong 
dependencies  of the normalized angular momentum transport on the microscopic viscosity and/or the basic rotation rate. In 
particular we do not see a decay of the angular momentum transport of the MRI for small magnetic Prandtl number in contrast 
to the conclusions of Fromang et al. (2007). As known from the bifurcation maps of the standard MRI the actual value 
of the magnetic Prandtl number is not very important if only the basic rotation and the magnetic background field are 
normalized without use of the microscopic viscosity.  Of course, one  can  split the Lundquist number in accordance to
\begin{equation}
{\rm S}= \frac{{\rm Rm}}{\sqrt{\beta}},
\label{Lund}
\end{equation}
with the plasma  $\beta=(\Om D \sqrt{\mu_0 \rho}/B_z)^2$. But in this case the dependence of $\alpha$ on ${\rm Rm}$ and 
$\beta$ such as  described by Longaretti \& Lesur (2010) is merely an artificial dependence. After our findings   the 
$\alpha_{\rm SS}$ only depends   on the linear product of the magnetic field, the electric conductivity and the disk 
size. The total value of $\alpha_{\rm SS}$ does not strongly exceed the limit of $10^{-3}$. In comparison with the 
results of  Longaretti \& Lesur the global simulations lead to values smaller by one order of magnitude.

%%%%%%%%%%%%%%%%%%%%%%%%%%%%%%%%%%%%%%%%%%
\section{Discussion}
%%%%%%%%%%%%%%%%%%%%%%%%%%%%%%%%%%%%%%%%%%%
The standard MRI is the instability of differential rotation under the presence of an axial magnetic field. Axisymmetric 
and nonaxisymmetric modes are excited for supercritical but not too strong fields. One must expect, however, that the 
nonaxisymmetric modes are quenched by too strong differential rotation. On the other hand, the nonaxisymmetric modes are 
important for any form of hydromagnetic dynamo and/or the evolution of small-scale turbulence and its angular momentum 
transport. It is this rotational quenching of the nonaxisymmetric modes excited close to the weak-field branch and its 
consequences for the angular momentum transport which is probed in the present paper.

We have shown by considering a dissipative cylindric TC flow that the MRI only forms nonaxisymmetric modes in a special 
domain in the Reynolds number-Lundquist number map. While for large $\rm Rm$ the minimum Lundquist number 
$\rm S_{\rm MIN}\simeq 1$ for the excitation of axisymmetric modes does not depend on $\rm Rm$ the minimum magnetic 
field for the excitation of nonaxisymmetric modes grows for growing $\rm Rm$. Both the weak-field limit and the 
strong-field limit of MRI have a positive slope for nonaxisymmetric modes (see  Kitchatinov \& R\"udiger 2010).
The inverse magnetic Mach number $\Omega_{\rm A} / \Omega_{\rm in}$ must exceed the value $\simeq 3\cdot 10^{-3}$ 
(for $\rm Pm=1$) in order to get nonaxisymmetric modes excited. This quenching effect is very similar for $\rm Pm<1$ 
but it is weaker for  $\rm Pm>1$ (see Eq. (\ref{seed1})). Fluids with  $\rm Pm>1$ seem to be favored, therefore, in 
numerical simulations.

For field amplitudes smaller than the given ones the standard MRI in cylinders is only formed by axisymmetric rolls 
which remained stable in our simulations with $\rm Rm$ up to 10$^3$. The domain of turbulence, therefore, proved to 
be (much) smaller than the entire domain of MRI. The most important numerical result of the simulations is that  for 
increasing magnetic Reynolds number the MRI develops from axisymmetric rolls to nonaxisymmetric `turbulence' and 
returns to axisymmetric rolls. We did not find sofar a second bifurcation of the roll instability  to more chaotic 
patterns (Fig. \ref{f71}).

The dependence of the wave numbers on the field strength is in accordance to the expectation that the axial scales grow 
for growing field. Figure \ref{f2} shows for $m>0$ how large values of $\rm S$ correspond to small values of $k$ as it follows 
from the Taylor-Proudman 
theorem. The same is true for $m=0$ but with a specialty:  close to the weak-field branch the axial wave length sinks 
down  to a minimum for increasing field amplitudes due to the influence of the dissipation. Beginning from this minimum 
for increasing field amplitudes the wave lengths grow in accordance to the Taylor-Proudman theorem (see Fig. \ref{wave}).

The angular momentum transport by the MRI has been computed with a nonlinear MHD spectral code which, of course,  is able 
to reproduce all the above given  results of the linear theory. In a first step only the angular momentum transport by 
the nonaxisymmetric modes (which in the model are the only one to be responsible for turbulence) has been computed. 
Expressed by the MRI-$\alpha$ (\ref{alfs}) the results are of a striking simplicity. We did not find an influence  of the 
numerical values of the magnetic Prandtl number and/or the magnetic Reynolds number on the resulting $\alpha_{\rm SS}$ 
(Fig. \ref{alpha_f}). The Lundquist number gives the only  influence  represented by the linear relation (\ref{alfs1}). 
The numerical value of the $\alpha_{\rm SS}$ taken at (say) $\rm S \simeq 100$ (1 Gauss for protoplanetary disks at R= 1 AU!) 
is only of order $10^{-3}$. This value strongly reminds on the numerical results of Brandenburg et al. (1996) obtained with 
shearing box simulations. Higher values of $\alpha_{\rm SS}$ require stronger fields. Nonlinear relations between 
$\alpha_{\rm SS}$ and the given axial field component  $\rm S$ have not been confirmed here.

The inclusion of the angular momentum transport by the axi\-sym\-metric `channel' modes gives only a slight modification 
of this picture. The maximum amplification of the angular momentum transport by the channel mode is characterized by a 
factor of two, which for slow rotation gives a stronger increase of the $\alpha_{\rm SS}$ than for fast rotation 
(Figs. \ref{alpha13} and \ref{alpha30}). The total angular momentum transport expressed by  $\alpha_{\rm SS}$ due to  
axisymmetric and nonaxisymmetric modes finally obtains a weak decrease for increasing Reynolds number. There is even  
indication that for large Reynolds number when the nonaxisymmetric modes eventually  become stable the transport by the 
axisymmetric modes yields nearly  the   $\alpha_{\rm SS}$   as the transport by the nonaxisymmetric modes for medium 
Reynolds numbers does.

%%%%%%%%%%%%%%%%%%%%%%%%%%%%%%%%%%%%%%%%%%%%%%%%%%%%%%%%%%%%%%%%%%%%%%%%%%%%%%%%%%%%%%%%%%%%%%%%%%%%%%%%%%%%%%%%%%%%%%%

\end{document}